\documentclass[aps,prd,nofootinbib]{revtex4}
\usepackage[colorlinks=true, pdfstartview=FitV, linkcolor=blue, citecolor=red, urlcolor=magenta]{hyperref}
\usepackage{graphicx}
\usepackage{latexsym}
\usepackage{amsmath}
\usepackage{amsfonts}
\usepackage{amssymb}
\usepackage{verbatim}
\usepackage{dsfont}

\newcommand{\be}{\begin{equation}}
\newcommand{\ee}{\end{equation}}
\newcommand{\bea}{\begin{eqnarray}}
\newcommand{\eea}{\end{eqnarray}}

\newcommand{\ben}{\begin{eqnarray}}
\newcommand{\een}{\end{eqnarray}}

\begin{document}

\title{Boundary effects on classical liquid density fluctuations}


\author{$^{1}$K. E. L. de Farias}
\email{klecio.limaf@gmail.com}

\author{ $^{2}$ Azadeh Mohammadi}
\email{azadeh.mohammadi@ufpe.br}

\author{$^{1}$ Herondy F. Santana Mota}
\email{hmota@fisica.ufpb.br}

\affiliation{$^{1}$Departamento de F\' isica, Universidade Federal da Para\' iba,\\  Caixa Postal 5008, Jo\~ ao Pessoa, Para\' iba, Brazil}
\affiliation{$^{2}$ Departamento de F\'isica, Universidade Federal de Pernambuco,\\
 Av. Prof. Moraes Rego, 1235, Recife - PE - 50670-901, Brazil.}



\begin{abstract}
In this paper, we study quantum vacuum fluctuation effects on the mass density of a classical liquid arising from the conical topology of an effective idealized cosmic string spacetime, as well as from the mixed, Dirichlet, and Neumann boundary conditions in Minkowski spacetime. In this context, we consider a phonon field representing quantum excitations of the liquid density, which obeys an effective Klein-Gordon equation with the sound velocity replaced by the light velocity. In the idealized cosmic string spacetime, the phonon field is subject to a quasi-periodic condition. Moreover, in Minkowski spacetime, the Dirichlet and Neumann boundary conditions are applied on one and also two parallel planes. We, thus, in each case, obtain closed analytic expressions for the two-point function and the renormalized mean-squared density fluctuation of the liquid. We point out specific characteristics of the latter by plotting their graphs.

\end{abstract}
 \maketitle


\section{Introduction}
Similar to photons as quantized light waves, phonons are quasi-particles that may be interpreted as quantized sound waves due to the atomic lattice's excitation. Hence, several properties can be examined in a phonon system related to relativistic quantum fields. This is particularly notable if the wavelengths of the quasi-particles are large compared with the interatomic distances, which means that the phonon dispersion relation is approximately linear. Although the term ``phonon" was given in 1931 by Igor Tamm, the phonon theory was developed by Einstein in 1907 \cite{einstein1907plancksche}, and Debye in 1912 \cite{debye1912theorie}. However, only in 1941, the phonon theory was applied to superfluid by Landau \cite{landau1941two} and posteriorly in a classical liquid context by Percus and Yevick (1958) \cite{percus1958analysis}. Subsequently, many studies of phonon properties in fluids were published as, for instance, superconduction by phonon-electron interaction \cite{axe1973influence}, phonon-phonon scattering \cite{maris1977phonon}, and in the heat capacity of several fluids \cite{bolmatov2012phonon}.

Furthermore, as it is well known, the presence of boundary conditions or spacetime with nontrivial topology may modify the quantum behavior of the system's properties. The case of a classical liquid is not different, that is, the local changes in its mass density depend on the effective spacetime geometry and the boundary condition under which the liquid is submitted. In the related context, Unruh \cite{unruh1981experimental} has proposed an experiment involving sound waves in a fluid as an analog model to study black hole evaporation. More recently, from a cosmological and astrophysical point of view, it was considered a phonon superfluid to study dark matter \cite{nicolis2011low,nicolis2018mutual,berezhiani2020effective}.

Due to similar properties, we can submit phonons to specific boundary conditions to obtain an analog Casimir effect \cite{casimir1948attraction}, as it is usual in the context of quantum field theory. Consequently, considering a liquid that simulates a background, it is possible to compute density fluctuations in the same way as done for the Casimir effect, although by replacing the speed of light with the speed of sound \cite{dzyaloshinskii1961general}. Originally, the density fluctuations were calculated by L. Ford \cite{ford2009fluid,ford2009phononic}, not only by considering nontrivial topology, including the one associated with an ideal cosmic string, and different boundary conditions but also by considering changes in the quantum states of the phonon. Cosmic string spacetime is a nontrivial and interesting topology proposed by Kibble \cite{kibble1976topology}, which is a line-like topological defect predicted in many extensions of the Standard Model of particle physics \cite{vilenkin2000cosmic,hindmarsh1995cosmic,allen1990evolution}.

In quantum field theory, phonons are represented by a real massless scalar field with spin$-0$. Since the scalar field may be submitted to boundary conditions, it codifies the variations in the quantum system, showing its different characteristics. This has been considered, for instance, by T. H. Boyer \cite{boyer1974van}, in the case of two parallel plates, where one of them was a perfect conductor plate, and the other an infinitely permeable one. To conduct this study, Boyer used mixed boundary condition to represent possible differences in the properties of the plates and found a repulsive result in which the force was multiplied by the factor $7/8$, in contrast with the attractive force present between the plates once they are submitted only to Dirichlet or Neumann boundary conditions, for instance. Hence, the difference in the multiplicative factors is attributed to plates with different properties.

The purpose of this paper is to generalize the result obtained in \cite{ford2009fluid,ford2009phononic} for the cosmic string by introducing the quasi-periodic condition given by $\Phi(t,r,\varphi,z)=e^{-2\pi i\beta}\Phi(t,r,\varphi+2\pi/q,z)$, therewith, the solution of the equation of motion will present the explicit dependence on the parameter $\beta$. The parameter $q$, on the other hand, encodes the conical structure of the spacetime \cite{vilenkin2000cosmic,hindmarsh1995cosmic,allen1990evolution}. In \cite{ford2009fluid,ford2009phononic}, the authors found the renormalized mean square density fluctuation for $q>0$, considering the particular case $\beta=0$, which represents periodic boundary condition. In this work, we obtain the renormalized mean square density fluctuation for the general case, $q>0$, and arbitrary $\beta$. Moreover, we study the influence of Dirichlet, Neumann, and mixed boundary conditions in the density fluctuations in the cases of one and two parallel planes in Minkowski spacetime.

The paper is organized as follows. In Sec.\ref{sec2} we give a brief overview of the phonon theory in a classical liquid. In Sec.\ref{sec3} we find a closed and exact analytical expression for the two-point function along with the mean square density fluctuation of the liquid, as a consequence of the imposition of a quasi-periodic condition on the massless scalar field whose modes propagates in the conical structure of a cosmic string, or disclination, spacetime. In Sec.\ref{sec4} we also explicitly calculate both the two-point function and mean square density fluctuation associated with the massless scalar field representing phonon modes of the liquid by imposing Dirichlet, Neumann, and mixed boundary conditions. Finally, in Sec.\ref{sec5} we present our conclusions.


\section{Phonon in a Liquid}
\label{sec2}

In this section, we give a brief overview of the quantum density fluctuation theory of a classical fluid, such as a liquid, coming from the zero-point oscillations, analogous to the one in relativistic quantum field theory. The phonon theory with phonons as quantized sound waves is built upon the presence of perturbations in the fluid mass density which can be written in the form $\rho'=\rho-\rho_0$, with $\rho_0$ being a constant mean mass density. Thereby, in order to find local density fluctuations in the phonon vacuum state, the perturbed mass density $\rho'$ may be related to a real massless scalar field $\phi$ according to the equation \cite{lifshitz2013statistical}
\begin{equation}
\frac{\partial\rho^{\prime}}{\partial t}=-\rho_0\nabla^2\phi.
\label{1.0}
\end{equation}
This is essentially the continuity equation for a liquid with velocity $\vec{v}\equiv \nabla\phi$ \cite{unruh1981experimental}, where the real scalar field is the velocity potential. By following the usual quantization rules, the quantum description for the liquid is reached once we replace the classical hydrodynamics quantities with operators expressed in terms of phonon annihilation and creation operators $\hat{c_k}$, $\hat{c}^{\dagger}_k$, satisfying the following commutation relation
\begin{equation}
\left[\hat{c}_k,\hat{c}^{\dagger}_{k^\prime}\right]=\delta_{kk^\prime},
\end{equation}
with $\delta_{kk^\prime}$ being either a Kronecker or a Dirac delta depending on whether the set of field modes $k$ is discrete or continuum, respectively.  As it has been said previously, in this work, we consider the quantization of sound waves in a fluid with a linear dispersion relation $\omega=u|k|$, with $u$ being the sound velocity in the liquid. This is a valid approximation as long as the wavelengths are much longer than the interatomic separation.

In the fluid theory, the density perturbation and velocity potential operators should also obey the following commutation rule
\begin{equation}
\hat{\phi}(\vec{r})\hat{\rho}^{\prime}(\vec{r}\;^{\prime})-\hat{\rho}^{\prime}(\vec{r}\;^{\prime})\hat{\phi}(\vec{r})=-i\hbar\delta^3(\vec{r}-\vec{r}\;^{\prime}),
\label{1.1}
\end{equation}
where $\delta^3(\vec{r}-\vec{r}\;^{\prime})$ is the Dirac delta function and \cite{lifshitz2013statistical}

\begin{equation}
\hat{\rho}^{\prime}(t,\vec{r})=-\frac{\rho_0}{u^2}\dot{\hat{\phi}}(t,\vec{r}).
\label{1.4}
\end{equation}
Clearly, replacing \eqref{1.4} into \eqref{1.0} gives the Klein-Gordon equation for a real massless scalar field. The only difference with the relativistic field theory is that one should consider the sound velocity, $u$, in the liquid instead of the light velocity, $c$, in a vacuum (for more details see \cite{lifshitz2013statistical}).

In the following, we study modifications on the quantum vacuum fluctuations by considering that the phonon modes propagate in the (3+1)-dimensional cosmic string spacetime under a quasi-periodic condition. We also consider the phonon modes propagating in Minkowski spacetime obeying mixed, Dirichlet, and Neumann boundary conditions. Note that the ideal cosmic string geometry can, for example, appear as a defect in a liquid crystal in the form of a disclination. We calculate the closed and analytical expressions for the two-point function and the renormalized mean square density fluctuation for each case.


\section{Phonons in the Cosmic String Spacetime}
\label{sec3}

The line element of an effective cosmic string or disclination spacetime considered in a liquid theory must contain the sound velocity $u$ replacing the light velocity $c$. In this sense, in $(3+1)-$dimensional effective cosmic string spacetime, we have, in cylindrical coordinates,
\begin{equation}
ds^2=g_{\mu\nu}dx^{\mu}dx^{\nu}=u^2dt^2-dr^2-r^2d\varphi^2-dz^2,
\label{2.0}
\end{equation}
where the spacetime coordinates are defined in the intervals: $r\geq0$, $\varphi\in[0,2\pi/q]$ and $t,z\in(-\infty,+\infty)$. As we have mentioned before, the parameter $q$ encodes the conical structure of the spacetime, which becomes a disclination parameter in condensed matter systems such as systems involving liquid crystals. In the latter case, the conical parameter may assume values $q>0$ \cite{katanaev1992theory}. When $q=1$, the conical structure disappears, and one recovers the Minkowski spacetime.

The Klein-Gordon equation in cylindrical coordinates, considering the line element \eqref{2.0}, is written as
\begin{equation}
\left[\frac{1}{u^2}\frac{\partial^2}{\partial t^2}-\frac{1}{r}\frac{\partial}{\partial r}\left(r\frac{\partial}{\partial r}\right)-\frac{1}{r^2}\frac{\partial^2}{\partial\varphi^2}-\frac{\partial^2}{\partial z^2}\right]\phi(t,r,\varphi,z)=0.
\label{2.6.1}
\end{equation}
Let us solve the above equation by submitting its solution to the quasi-periodic condition
\begin{equation}
\phi(t,r,\varphi,z)=e^{-2\pi i\beta}\phi(t,r,\varphi+2\pi/q,z),
\label{2.1}
\end{equation}
with $0\leq\beta<1$. Thus, for a real massless scalar field in the effective cosmic string spacetime characterized by the line element (\ref{2.0}), the solution of Eq. \eqref{2.6.1}, under the quasi-periodic condition \eqref{2.1}, is given by \cite{klecio2020quantum}
\begin{equation}
\phi(t,r,\varphi,z)=Ae^{-i\omega_k t}e^{i\nu z}e^{iq(n+\beta)\varphi}J_{q|n+\beta|}(\eta r).
\label{2.2}
\end{equation}
The parameter $A$ in the above solution is a normalization constant, $\omega^2_k=u^2(\nu^2+\eta^2)$ is the dispersion relation, $k=(n,\eta,\nu)$ is the set of quantum numbers and $J_{\mu}(x)$ is the Bessel function of first kind.
The normalization constant $A$ can be obtained by considering the commutation rule \eqref{1.1} as well as \eqref{1.4} which leads to
\begin{equation}
|A_k|=\sqrt{\frac{qu\hbar\eta}{8\pi^2\rho_0\omega_k}}.
\label{2.7}
\end{equation}

As it is known, phonons can be viewed as sound wave excitations of a real massless scalar field, in our case, the complete normalized solution \eqref{2.2}, with \eqref{2.7}. These excitations are formally constructed once we quantize the real massless scalar field in terms of phonon annihilation and creation operators $\hat{c}_k$ and $\hat{c}^{\dagger}_k$, respectively. Thereby, the field operator is written as
\begin{equation}
\hat{\phi}(t,r,\varphi,z)=\sum_{\{k\}}\left[A_k\hat{c}_ke^{-i\omega_k t+i\nu z+iq(n+\beta)\varphi}+A_k^{*}\hat{c}^{\dagger}_ke^{i\omega_k t-i\nu z-iq(n+\beta)\varphi}\right]J_{q|n+\beta|}(\eta r),
\label{2.5}
\end{equation}
where
\begin{equation}
\sum_{\{k\}}=\int_{\infty}^{\infty}\!\!\!\!d\nu\int^{\infty}_{0}\!\!\!\!d\eta\sum_{n=-\infty}^{\infty},
\label{symbol}
\end{equation}
is the sum over all quantum numbers.

Let us now turn to the calculation of the two-point function $G(w,w^{\prime})$, where $w \equiv (t,r,\varphi,z)$. The two-point function is important to calculating the renormalized mean square density fluctuation, which is the physical observable of interest in our investigation. We can then make use of the field operator (\ref{2.5}) to calculate $G(w,w^{\prime})$ according to \cite{birrell1984quantum}
\begin{eqnarray}
G(w, w^{\prime})&=&\langle\hat{\phi}(w)\hat{\phi}(w^{\prime})\rangle,
\label{2.9}
\end{eqnarray}
where $\langle\;\; \rangle$ is vacuum expectation value. The substitution of Eq. (\ref{2.5}) into Eq. \eqref{2.9} provides
\begin{eqnarray}
G(w, w^{\prime})=\sum_{\{k\}}\frac{qu\hbar\eta}{8\pi^2\rho_0\omega_k}e^{i\nu\Delta z+iq(n+\beta)\Delta\varphi-i\omega_k\Delta t}J_{q|n+\beta|}(r\eta)J_{q|n+\beta|}(r^{\prime}\eta),
\label{2.91}
\end{eqnarray}
with $\Delta t = t - t'$, $\Delta\varphi = \varphi- \varphi'$ and $\Delta z = z- z'$. Furthermore, in order to solve the integrals in $\nu$ and $\eta$ present in the two-point function \eqref{2.91} we can make the Wick rotation $i\Delta t=\Delta\tau$ and use the following identity
\begin{equation}
\frac{e^{-\omega_k\Delta\tau}}{\omega_k}=\frac{2}{\sqrt{\pi}}\int_{0}^{\infty}ds e^{-s^2\omega_k^2-\Delta\tau^2/4s^2}.
\label{2.10}
\end{equation}
The integral in $\eta$ can, then, be obtained by making use of Eq. (21) from Ref. \cite{bragancca2019vacuum}. Thus, the two-point function \eqref{2.91} becomes
\begin{eqnarray}
G(w,w^{\prime})&=&\frac{qu\hbar}{8\pi^2\rho_0}\frac{e^{iq\beta\Delta\varphi}}{\sqrt{\pi}}\int_{-\infty}^{\infty}d\nu e^{i\nu\Delta z}\int_{0}^{\infty}\frac{ds}{s^2}e^{-s^2\nu^2-\frac{\Delta\zeta^2}{4s^2}}\sum_{n=-\infty}^{\infty}e^{inq\Delta\varphi}I_{q|n+\beta|}(rr^{\prime}/2s^2),
\label{2.10}
\end{eqnarray}
where $\Delta\zeta^2=\Delta\tau^2+r^2+r^{\prime2}$. Finally, we can observe that there are still integrals in $s$ and $\nu$ and a sum in $n$ to be performed. The integral in $\nu$ is a gaussian-type integral and can easily be obtained. Moreover, the sum in $n$ can be worked out by using the summation formula (25) from Ref. \cite{bragancca2019vacuum}. Subsequently, the integral in $s$ can be solved which provides the closed and exact form for the two-point function, i.e.,
\begin{eqnarray}
G(w,w^{\prime})&=&\frac{qu\hbar e^{iq\beta\Delta\varphi}}{8\pi^2\rho_0rr^{\prime}}\left\{\frac{1}{q}\sum_n e^{i\beta(2\pi n-q\Delta\varphi)}\frac{1}{\sigma_n}\right.\nonumber\\
\ &\ &\left.-\frac{1}{2\pi i}\sum_{j=+,-}je^{ijq\beta\pi}\int_{0}^{\infty}dy\frac{1}{\sigma_y}\frac{\cosh[qy(1-\beta)]-\cosh(q\beta y) e^{-iq(j\pi+\Delta\varphi)}}{\cosh(qy)-\cos[q(\Delta\varphi+j\pi)]}\right\},
\label{2.11}
\end{eqnarray}
where
\begin{eqnarray}
\sigma_n&=&\frac{\Delta\zeta^2}{rr^{\prime}}+\frac{\Delta z^2}{2rr^{\prime}}-\cos(2\pi n/q-\Delta\varphi),\nonumber\\
\sigma_y&=&\frac{\Delta\zeta^2}{rr^{\prime}}+\frac{\Delta z^2}{2rr^{\prime}}+\cosh y.
\label{inter}
\end{eqnarray}
Note that the new sum in $n$ in the two-point function, \eqref{2.11}, is restricted to the interval \cite{de2015vacuum, bragancca2019vacuum, klecio2020quantum}
\begin{equation}
-\frac{q}{2}+\frac{\Delta\varphi}{\varphi_0}\leq n\leq\frac{q}{2}+\frac{\Delta\varphi}{\varphi_0}.
\end{equation}
One should also note that for $q<2$, the only contribution in the first term on the r.h.s of Eq. \eqref{2.11} comes from the term $n=0$, which is the Minkowski contribution to the two-point function. As it is widely known, the latter diverges in the coincidence limit $w^{\prime}\to w$, and it should be subtracted to calculate the physical observables, thereby providing a finite renormalized quantity. As a consequence, the renormalized two-point function is obtained as
\begin{equation}
G_{\text{ren}}(w,w') = G(w,w') - G_{\text{M}}(w,w'),
\label{CL}
\end{equation}
where $G_{\text{M}}(w,w')$ is the $n=0$ term of the sum in \eqref{2.11} representing the Minkowski contribution, that is,
\begin{equation}
G_{\text{M}}(w,w') =\frac{u\hbar }{8\pi^2\rho_0rr^{\prime}}\frac{1}{\sigma_0},
\label{MC}
\end{equation}
where $\sigma_0$ is given by Eq. \eqref{inter}. We can see that the Minkowski contribution above is clearly divergent in the coincidence limit $w'\rightarrow w$. We should point out that except for the constants associated with the phonon system, the two-point function in Eq. \eqref{2.11} is similar to the Wightman function in Ref. \cite{klecio2020quantum} for the massless case, showing the consistency of our result.
%
\subsection{Mean square density fluctuation}

\begin{figure}[!htb]
\begin{center}
\includegraphics[width=0.4\textwidth]{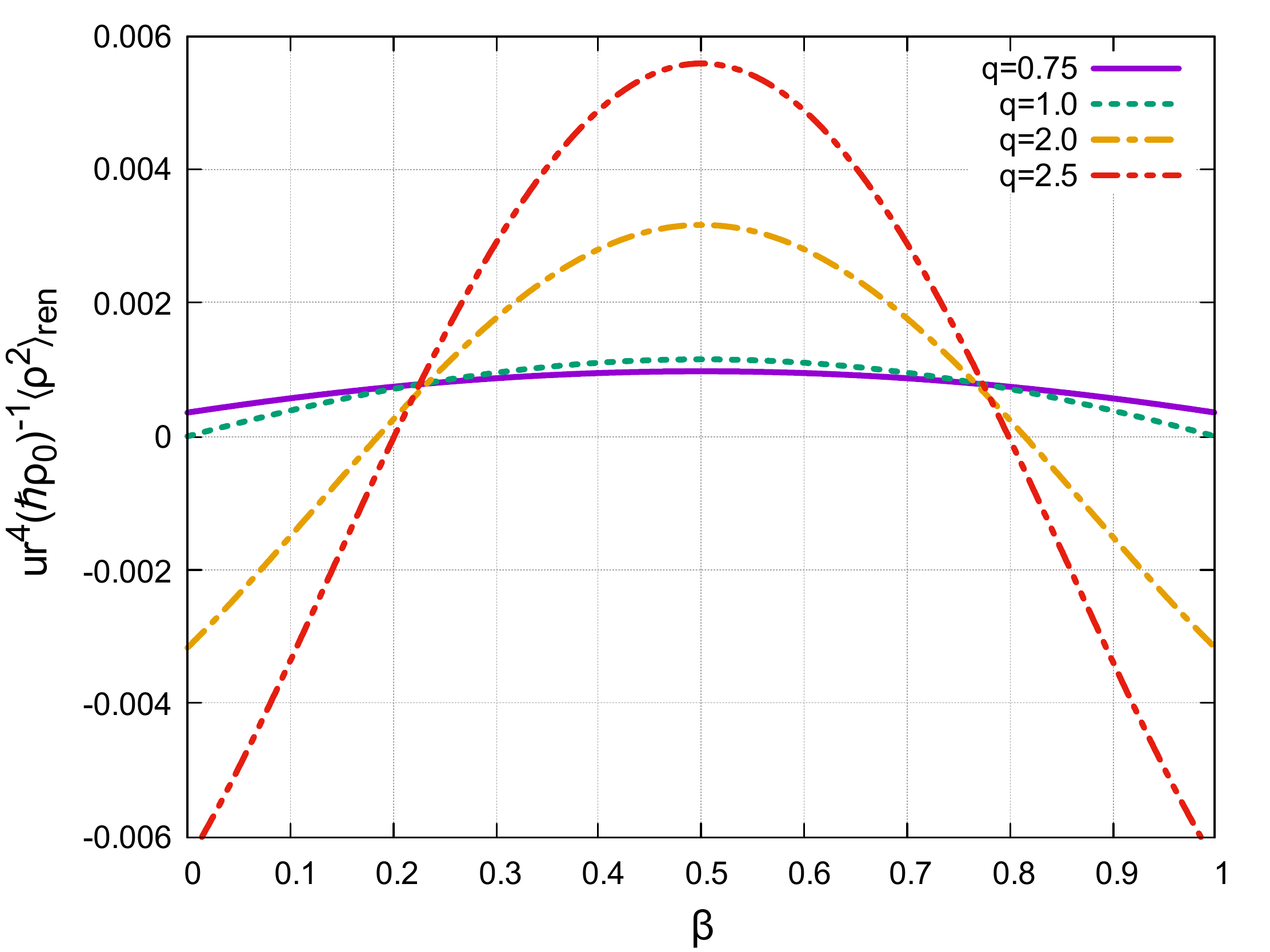}
\caption{Dimensionless renormalized mean squared density fluctuation, \eqref{2.16}, in terms of $\beta$, considering different values for the cosmic string parameter $q$.\small}
\label{figure1}
\end{center}
\end{figure}

The local effect on the mean square density fluctuation due to the quasi-periodic condition \eqref{2.1} in a cosmic string spacetime \eqref{2.0} can be computed by using Eq. \eqref{1.4}. The vacuum expectation value $\langle \hat{\rho}(w) \rangle$ vanishes due to the fact that the operator $\hat{\rho}(w)$ is linear in the annihilation and creation operators $\hat{c}^{\dagger}_k$ and $\hat{c}_k$. To compute the mean square density fluctuation, we have to consider the product
\begin{eqnarray}
\hat{\rho}(w)\hat{\rho}(w')&=&\frac{\rho_0^2}{u^2}\frac{\partial^2}{\partial t\partial t^{\prime}}[\hat{\phi}(w)\hat{\phi}(w')].
\label{2.14}
\end{eqnarray}
Thus, taking the vacuum expectation value of the above expression, one can find the two-point function mean squared density fluctuation, in terms of $G(w,w')$, in the form
\begin{eqnarray}
\langle\hat{\rho}(w)\hat{\rho}(w')\rangle&=&\frac{\rho_0^2}{u^2}\frac{\partial^2}{\partial t\partial t^{\prime}}G(w,w^{\prime}).
\label{2.15}
\end{eqnarray}
Now, by taking the coincidence limit $w^{\prime}\rightarrow w$ and subtracting the Minkowski contribution \eqref{MC}, we can finally find a closed form for the renormalized mean squared density fluctuation
\begin{eqnarray}
\langle\rho^2\rangle_{\rm ren}&=&\frac{\rho_0^2}{u^2}\lim_{w^{\prime}\rightarrow w}\frac{\partial^2}{\partial t\partial t^{\prime}}G_{\rm ren}(w,w^{\prime})\nonumber\\
\ &=&-\frac{\hbar\rho_0}{32\pi^2ur^4}\left\{2\sum_{n=1}^{[q/2]}\!^* \frac{\cos(2\beta\pi n)}{\sin^4(\pi n/q)}-\frac{q}{\pi}\int_0^{\infty}dy\frac{M(y,\beta,q)}{\cosh^4(y/2)}\right\},
\label{2.16}
\end{eqnarray}
where the function $M(y,\beta,q)$ is defined as
\begin{equation}
M(y,\beta,q)=\frac{\cosh(q\beta y) \sin[q\pi(1-\beta)]+\cosh[qy(1-\beta)]\sin(q\beta\pi)}{\cosh(qy)-\cos(q\pi)}.
\end{equation}
It is worth mentioning that $[q/2]$ in the sum present in the renormalized squared density fluctuation \eqref{2.16}, represents the integer part of $q/2$, and the sign $(*)$ means that in the case of an integer $q$, the sum in $n$ must be replaced by
\begin{equation}
\sum_{n=1}^{[q/2]}\rightarrow\frac{1}{2}\sum_{n=1}^{q-1}.
\label{sumC}
\end{equation}
One should note that, the mean squared density fluctuation is an elementary function of $r$, that is, $\langle\rho^2\rangle_{\rm ren}\propto r^{-4}$, which means the expression \eqref{2.16} diverges when $r\rightarrow 0$ and goes to zero when $r\rightarrow\infty$. In the absence of the cosmic string, i.e., $q=1$, the only contribution is due to the quasi-periodic condition \eqref{2.1}, originating from the second term on the r.h.s of Eq. \eqref{2.16}. On the other hand, in the absence of the quasi-periodic condition, i.e., $\beta=0$, the effect on the mean squared density fluctuation is entirely due to the nontrivial topology of the cosmic string. In Fig.~\ref{figure1}, we have plotted the mean squared density fluctuation presented in Eq. \eqref{2.16}, as a function of the parameter $\beta$ for several values of the cosmic string parameter $q$. Notice that, if we set $q=1$ and $\beta=0$ in Eq. \eqref{2.16}, i.e., in the absence of any boundary condition, the renormalized mean squared density fluctuation vanishes, as it should.
%
%
\subsection{Particular Cases}
Let us now turn our attention to three particular cases for the mean squared density fluctuation obtained in Eq. \eqref{2.16}, namely, the case $\beta=0$ (only cosmic string), $q=1$ (only periodic-condition), and the case $\beta=\frac{1}{2}$.

For a purely cosmic string conical topology contribution, $\beta=0$, from Eq. \eqref{2.16}, we have
\begin{equation}
\langle\rho^2\rangle_{\text{ren}}=-\frac{\hbar\rho_0}{32\pi^2ur^4}\left\{2\sum_{n=1}^{[q/2]}\!^* \frac{1}{\sin^4(\pi n/q)}-\frac{q}{\pi}\int_0^{\infty}\!\!dy\ \frac{M(y,0,q)}{\cosh^4(y/2)}\right\}.
\label{2.18}
\end{equation}
Furthermore, if we consider only integer values of $q$ in Eq. \eqref{2.18}, the integral term vanishes, and the only contribution comes from the first term on the r.h.s. Thus, by using Eq. \eqref{sumC}, we are able to perform the sum in $n$ to obtain
\begin{equation}
\langle\rho^2\rangle_{\text{ren}}=-\frac{\hbar\rho_0}{1440\pi^2ur^4}\ (-11+10q^2+q^4),
\label{2.18.1}
\end{equation}
which is always negative for any value of $q$. In fact, the above expression for the mean square density fluctuation is an analytic function of $q$ and, consequently, can be extended to any value of $q$, besides the integer values. This can be verified by numerically checking that both expressions \eqref{2.18} and \eqref{2.18.1} provide the same result for any value of $q$. Note that Eq. \eqref{2.18.1} vanishes for $q=1$, as expected since the cosmic string topology disappears. Note also that Eq. \eqref{2.18.1} is consistent with the result found in Ref. \cite{ford2009fluid}.

The case where we have only fluctuation effects due to the quasi-periodic condition \eqref{2.1}, by taking $q=1$ in Eq. \eqref{2.16}, only the second term on the r.h.s gives contribution, that is,
\begin{eqnarray}
\langle\rho^2\rangle_{\text{ren}}&=&\frac{\hbar\rho_0}{64\pi^3 ur^4}\int_0^{\infty}\!\!dy\frac{\sin(\beta\pi)\left(\cosh(\beta y)+\cosh[y(1-\beta)]\right)}{\cosh^6(y/2)}\nonumber\\
\ &=&\frac{\hbar\rho_0}{48\pi^2ur^4}\beta(\beta^2-1)(\beta-2),
\label{2.19}
\end{eqnarray}
where the integral in $y$ has been exactly solved. The result above, for the nonzero mean squared density fluctuation as a consequence of the quasi-periodic condition is exact and it vanishes for $\beta=0$.

Finally, we consider the case where $\beta=1/2$, i.e., the twisted scalar field. From Eq. \eqref{2.16} we, then, have
\begin{equation}
\langle\rho^2\rangle_{\text{ren}}=-\frac{\hbar\rho_0}{32\pi^2r^4u}\left\{\sum_{n=1}^{[q/2]}\!^*\frac{2\cos(n\pi)}{\sin^4(\pi n/q)}-\frac{q}{\pi}\int_0^{\infty}\!\!dy\frac{M(y,1/2,q)}{\cosh^4(y/2)}\right\}.
\label{2.17}
\end{equation}
Furthermore, let us take only integer values of the cosmic string parameter $q$. In this case, $\mathcal{M}(y,1/2,q)$=0, and as a consequence, only the first term on the r.h.s of Eq. \eqref{2.17} contributes to the mean squared density fluctuation. This gives
\begin{eqnarray}
\langle\rho^2\rangle_{\text{ren}}&=&-\frac{\hbar\rho_0}{16\pi^2ur^4}\sum_{n=1}^{q-1}\frac{(-1)^n}{\sin^4(\pi n/q)}\nonumber\\
&=&\frac{\hbar\rho_0}{11520\pi^2ur^4}(88+40q^2+7q^4),
\label{2.17.1}
\end{eqnarray}
where we have used Eq. \eqref{sumC}. Although we have obtained the above result considering integer values of $q$, Eq. \eqref{2.17.1} is an analytic function of $q$ and can be extended for all values of the cosmic string parameter. In fact, a numerical check shows that Eqs. \eqref{2.17} and \eqref{2.17.1} give the same result for any value of $q$. Moreover, as it can be seen, the expression \eqref{2.17.1} is always positive for phonons in a conical spacetime. Note that, by comparing Eq. \eqref{2.19}, for $\beta=1/2$, with Eq. \eqref{2.17.1}, for $q=1$, one can also see that they provide the same result, once again showing the consistency of our results. In the context of liquids, and considering phonons excitations, the results in Eqs. \eqref{2.16}, \eqref{2.19} and \eqref{2.17.1} have been obtained for the first time here, to the best of our knowledge.

\section{Dirichlet, Neumann and mixed boundary conditions }
\label{sec4}
%
In this section, we concentrate on the study of phonon modes subjected to Dirichlet, Neumann and mixed boundary condition in Minkowski spacetime. These boundary conditions will be taken by considering one and two parallel planes, consequently showing modifications on the mean squared density fluctuations of the liquid.

\subsection{One plane}
Let us start by considering Dirichlet and Neumann boundary conditions applied on a plane placed at the point $z=0$. Thereby, the normalized scalar field operator solution to the Klein-Gordon equation, $\Box\hat{\phi}(x)=0$, under Dirichlet and Neumann boundary conditions, is straightforward and is given by
\begin{eqnarray}
\varphi(w)=\sum_{\{k\}} \left(\frac{\hbar u}{(2\pi)^3\omega_k\rho_0}\right)^{\frac{1}{2}}\left(c_{k}e^{i\omega_k t-ik_xx-ik_yy}+c^{\dagger}_{k}e^{-i\omega_k t+ik_xx+ik_yy}\right)
\left\{
\begin{array}{c}
\sin(k_zz) \\
\cos(k_zz)
\end{array}
\right\},
\end{eqnarray}
where $w$ stands for the flat spacetime cartesian coordinates $(x,y,z)$, $\{k\}$ stand for the set of continuum quantum numbers $(k_x,k_y,k_z)$, $\omega_k^2=u^2(k_x^2 +k_y^2+k_z^2)$ is the dispersion relation, and the functions $\sin(nz)$ and $\cos(nz)$ represent the Dirichlet and Neumann boundary conditions, respectively. The two-point function in these cases are found to be
\begin{equation}
G(w,w^{\prime})=\frac{u\hbar}{4\pi^2\rho_0}\left[\frac{1}{\Delta z^2-\Delta\zeta^2}\pm\frac{1}{(z+z^{\prime})^2+\Delta\zeta^2}\right],
\label{3.01}
\end{equation}
with $\Delta\zeta^2\equiv \Delta x^2+\Delta y^2-\Delta t^2$. Note that the minus and plus signs stand for Dirichlet and Neumann boundary condition solutions, respectively. The first term on the r.h.s is the divergent Minkowski contribution in the coincidence limit $w'\rightarrow w$ and should be subtracted in the renormalization process. Hence, from Eqs. \eqref{2.15} and \eqref{3.01}, the renormalized mean squared density fluctuation is given by
\begin{equation}
\langle\rho^2\rangle_{\rm ren}=\pm\frac{\hbar\rho_0}{32u\pi^2}\frac{1}{z^4},
\label{3.02}
\end{equation}
where $z$ is the distance from the boundary. Note that the above result, for Dirichlet and Neumann boundary conditions, diverges on the boundary plane. In particular, the expression for Neumann boundary condition is consistent with the result obtained in Refs. \cite{ford2009fluid,ford2009phononic}.
%


\subsection{Two parallel planes}

Now, we wish to consider the phonon scalar field modes subject to mixed, Dirichlet, and Neumann boundary conditions on two parallel planes at $z=0$ and $z=a$. As expected, these boundary conditions make the quantum vacuum fluctuations of the phonon scalar field to be modified, resulting in a nonzero mean squared density fluctuation, as we shall see below.

\subsubsection{Mixed Boundary Condition}
In order to analyze the mixed boundary condition effects on the phonon modes, we consider two parallel planes at the points $z=0$ and $z=a$, where we require the scalar field to obey Dirichlet boundary condition  in the former point and obey the Neumann boundary condition in the latter one, i.e.,
\begin{equation}
\phi(t,x,y,z=0)=0,\qquad\qquad\qquad \partial_z \phi(t,x,y,z)|_{z=a}=0.
\label{3.1}
\end{equation}
The normalized scalar field operator solution to the Klein-Gordon equation, under the mixed boundary condition above, is written in terms of the phonon annihilation and creation operators, $c_k$ and $c_k^{\dagger}$, in the form
\begin{equation}
\hat{\phi}(w)=\sum_{\{k\}} \left(\frac{\hbar u}{8a\pi^2\omega_k\rho_0}\right)^{\frac{1}{2}}\left(c_{k}e^{-i\omega_k t+ik_xx+ik_yy}+c^{\dagger}_{k}e^{i\omega_k t-ik_xx-k_yy}\right)\sin(k_nz),
\label{3.5}
\end{equation}
where the momentum in the $z$-direction has been discretized, that is, $k_n=\frac{(2n+1)\pi}{2a}$, with $n= 1, 2, 3, ...$\; \cite{saharian2007generalized}. Note also that $\{k\}$ stands for the quantum numbers $(k_x,k_y,n)$, and $\omega_k^2 =u^2(k_x^2 +k_y^2+k_n^2)$ is the dispersion relation.

As before, in order to obtain the mean squared density fluctuation, we need to find the two-point function. Therefore, by using the scalar field solution operator, \eqref{3.5}, in Eq. \eqref{2.9},  for the two-point function, we have
\begin{eqnarray}
G(w,w^{\prime})&=&\frac{\hbar u}{4a\pi^2\rho_0}\int dk_xdk_ye^{-ikr\cos\theta}\sum_{n=0}^{\infty} \frac{e^{i\omega_n\Delta t}}{\omega_n}\sin(k_nz)\sin(k_nz^{\prime})\nonumber\\
\ &=&\frac{\hbar u}{2a\pi\rho_0}\int_0^{\infty} dk \;k \; J_0(kr)\sum_{n=0}^{\infty} \frac{e^{i\omega_n\Delta t}}{\omega_n}\sin(k_nz)\sin(k_nz^{\prime}),
\label{3.6}
\end{eqnarray}
where $r=\sqrt{(x-x^{\prime})^2+(y-y^{\prime})^2}$. To obtain the result in Eq.\eqref{3.6}, we have used polar coordinates for $k_x$ and $k_y$, and solved the angular integral in $\theta$ which gave rise to the Bessel function $J_0(kr)$ of the first kind. Note that, in this equation there remain still both the integral in $k$ and the summation in $n$ to be solved. The latter can be worked out by applying the following Abel-Plana formula \cite{saharian2007generalized}
\begin{equation}
\sum_{n=0}^{\infty}f\left(n+\frac{1}{2}\right)=\int_0^{\infty}\;dx\; f(x)-i\int_0^{\infty}\;dx \; \frac{f(ix)-f(-ix)}{e^{2\pi x}+1} .
\label{3.6.1}
\end{equation}
With this method, the divergent Minkowski contribution becomes evident, and we can renormalize the two-point function by removing this term.
To follow the procedure, we substitute Eq. \eqref{3.6} in the above Abel-Plana formula, resulting in
\begin{eqnarray}
G(w,w^{\prime})&=&\frac{\hbar u}{2a\pi\rho_0}\int_0^{\infty} dk \;k\;  J_0(kr)\left\{ \int_0^{\infty}\;dx\; \frac{e^{i\omega_x\Delta t}}{\omega_x}\sin(\alpha z)\sin(\alpha z^{\prime}) \right.\nonumber\\
&-&\left.2\int_{ka/\pi}^{\infty} dx \;  \frac{\cosh(\Delta\sqrt{\alpha^2-k^2})}{e^{2\pi x}+1}\;  \frac{\sin(i\alpha z)\sin(i\alpha z^{\prime})}{\sqrt{\alpha^2-k^2}}  \right\},
\label{3.7}
\end{eqnarray}
where $\omega_x=\sqrt{k^2+\alpha^2}$, and $\alpha=\frac{x\pi}{a}$. The expression above for the two-point function includes the divergent Minkowski contribution coming from the first term on the r.h.s, as well as the finite contribution in the second term. Thus, the renormalized two-point function can be written as
\begin{equation}
G_{\text{ren}}(w,w^{\prime}) \equiv G(w,w^{\prime})-G_{\text{M}}(w,w^{\prime}),
\label{ren}
\end{equation}
with $G_{\text{M}}(w,w^{\prime})$ being the divergent Minkowski contribution, which we shall calculate below. For this purpose, let us consider the first term on the r.h.s of Eq. \eqref{3.7}, i.e.,
\begin{eqnarray}
G_1(w,w^{\prime})&=&\frac{\hbar u}{2\pi a\rho_0} \int_0^{\infty}\;dx \sin(\alpha z)\sin(\alpha z^{\prime})\int_0^{\infty}\; dk\;k J_0(kr)\; \frac{e^{i\Delta t\sqrt{k^2+\alpha^2}}}{\sqrt{k^2+\alpha^2}}.
\label{3.8}
\end{eqnarray}
By using the Euler's formula, the integral in $k$ can be solved (see Ref. \cite{prudnikov1986integrals}\footnote{pg. 203, section 2.12.23, Eq. (8).}). This provides
\begin{eqnarray}
G_1(w,w^{\prime})&=&\frac{\hbar u}{2\pi^2 a\rho_0} \int_0^{\infty}\;dx\;  \sin(\alpha z)\sin(\alpha z^{\prime})\; \frac{e^{-\alpha\sqrt{r^2-\Delta t^2}}}{\sqrt{r^2-\Delta t^2}}\nonumber\\
\ &=&\frac{\hbar u}{4\pi^2 \rho_0} \int_0^{\infty}\;d\alpha \; [\cos(\alpha(z-z^{\prime}))-\cos(\alpha(z+z^{\prime}))]\; \frac{e^{-\alpha\sqrt{r^2-\Delta t^2}}}{\sqrt{r^2-\Delta t^2}}\nonumber\\
\ &=&\frac{\hbar u}{4\pi^2\rho_0}\left\{\frac{1}{\Delta z^2+\Delta\zeta^2} - \frac{1}{(z+z^{\prime})^2+\Delta\zeta^2}\right\},
\label{3.9}
\end{eqnarray}
where $\Delta\zeta^2=r^2-\Delta t^2$. The first term on the r.h.s of Eq \eqref{3.9} is the Minkowski contribution, $G_{\text{M}}(w,w^{\prime})$, which should be removed since it diverges when we take the coincidence limit $w^{\prime}\to w$. One should note that, the second term is finite in the coincidence limit, and is the contribution due to the presence of the Dirichlet boundary condition applied on the first plane.

Now, we can proceed with the second term on the r.h.s of Eq. \eqref{3.7}, that is,
\begin{eqnarray}
G_2(w,w^{\prime})&=&\frac{\hbar u}{\pi\rho_0a}\int_{0}^{\infty}dk\;kJ_0(kr)\int_{ka/\pi}^{\infty}dx\; \frac{\cosh(\Delta t\sqrt{\alpha^2-k^2})}{e^{2\pi x}+1} \; \frac{\sin(i\alpha z)\sin(i\alpha z^{\prime})}{\sqrt{\alpha^2-k^2}}\nonumber\\
\ &=&\frac{\hbar u}{\pi\rho_0a}\int_{0}^{\infty}dx\;\frac{\sin(i\alpha z)\sin(i\alpha z^{\prime})}{e^{2\pi x}+1}\int_{0}^{\alpha}dk\; kJ_0(kr) \; \frac{\cos(i\Delta t\sqrt{\alpha^2-k^2})}{\sqrt{\alpha^2-k^2}}.
\label{3.10}
\end{eqnarray}
The integral in $k$ in the above two-point function contribution can be solved employing the same method as in Ref. \cite{prudnikov1986integrals}\footnote{pg. 201, section 2.12.21, Eq. (6).}. Hence, it results in the following expression
\begin{eqnarray}
G_2(w,w^{\prime})=\frac{\hbar u}{\pi^2\rho_0}\int_{0}^{\infty}d\alpha \; \frac{\sin(i\alpha z)\sin(i\alpha z^{\prime})}{e^{2a\alpha}+1}\;  \frac{\sin(\alpha\sqrt{r^2-\Delta t^2})}{\sqrt{r^2-\Delta t^2}}.
\label{3.12}
\end{eqnarray}

Finally, from Eq. \eqref{ren}, the renormalized two-point function is found by collecting the second term on the r.h.s of Eq. \eqref{3.9}, along with the expression in Eq. \eqref{3.12}. This leads to
\begin{eqnarray}
G_{\rm ren}(w,w^{\prime})=-\frac{\hbar u}{\pi^2\rho_0}\left\{\frac{1}{4}\frac{1}{\left[(z+z^{\prime})^2+\Delta\zeta^2\right]} - \int_{0}^{\infty}d\alpha\ \frac{1}{e^{2a\alpha}+1} \ \frac{\sin(\alpha\sqrt{r^2-\Delta t^2})}{\sqrt{r^2-\Delta t^2}} \sin(i\alpha z)\sin(i\alpha z^{\prime}) \right\}.
\label{3.12.1}
\end{eqnarray}

For completeness, let us also consider both parallel planes subject to either Dirichlet or Neumann boundary condition. In this case, the scalar field operator solution to the Klein-Gordon equation is given by
\begin{eqnarray}
\varphi(w)=\sum_{\{k\}} \left(\frac{\hbar u}{4\pi^3\omega_k\rho_0}\right)^{\frac{1}{2}}\left(c_{k}e^{i\omega_k t-ik_xx-ik_yy}+c^{\dagger}_{k}e^{-i\omega_k t+ik_xx+ik_yy}\right)
\left\{
\begin{array}{c}
\sin{\frac{n\pi}{a}z} \\
\cos{\frac{n\pi}{a}z}
\end{array}
\right\},
\label{3.03}
\end{eqnarray}
where $\sin{\frac{n\pi}{a}z}$ $(n=1, 2, 3, \dots)$ indicates the solution under the Dirichlet boundary condition and $\cos{\frac{n\pi}{a}z}$ $(n=0, 1, 2, 3,\dots)$ under the Neumann boundary condition \cite{saharian2007generalized}. Hence, by following the same steps shown above, for the mixed boundary condition case, the result for either Dirichlet or Neumann applied on the two parallel planes is given by
\begin{eqnarray}
G_{\rm ren}(x,x^{\prime})=\frac{\hbar u}{\pi^2\rho_0}\left\{\mp\frac{1}{4}\frac{1}{\left[(z+z^{\prime})^2+\Delta\zeta^2\right]} + \int_{0}^{\infty}d\alpha\ \frac{1}{e^{2a\alpha}-1}\  \frac{\sin(\alpha\sqrt{r^2-\Delta t^2})}{\sqrt{r^2-\Delta t^2}}\left[
\begin{array}{c}
\sin(i\alpha z)\sin(i\alpha z^{\prime}) \\
\cos(i\alpha z)\cos(i\alpha z^{\prime})
\end{array}
\right]
\right\},
\label{3.04.1}
\end{eqnarray}
where we have used the Abel-Plana formula presented in Ref. \cite{saharian2007generalized} (see Eq. (2.16)). In what follows, we shall use the two-point function in Eq. \eqref{3.12.1} for the mixed boundary condition case, and  in Eq. \eqref{3.04.1} for either Dirichlet or Neumann boundary conditions, to calculate the mean squared density fluctuation.

\subsubsection{Mean Square Density}

\begin{figure}[!htb]
\begin{center}
\includegraphics[width=0.5\textwidth]{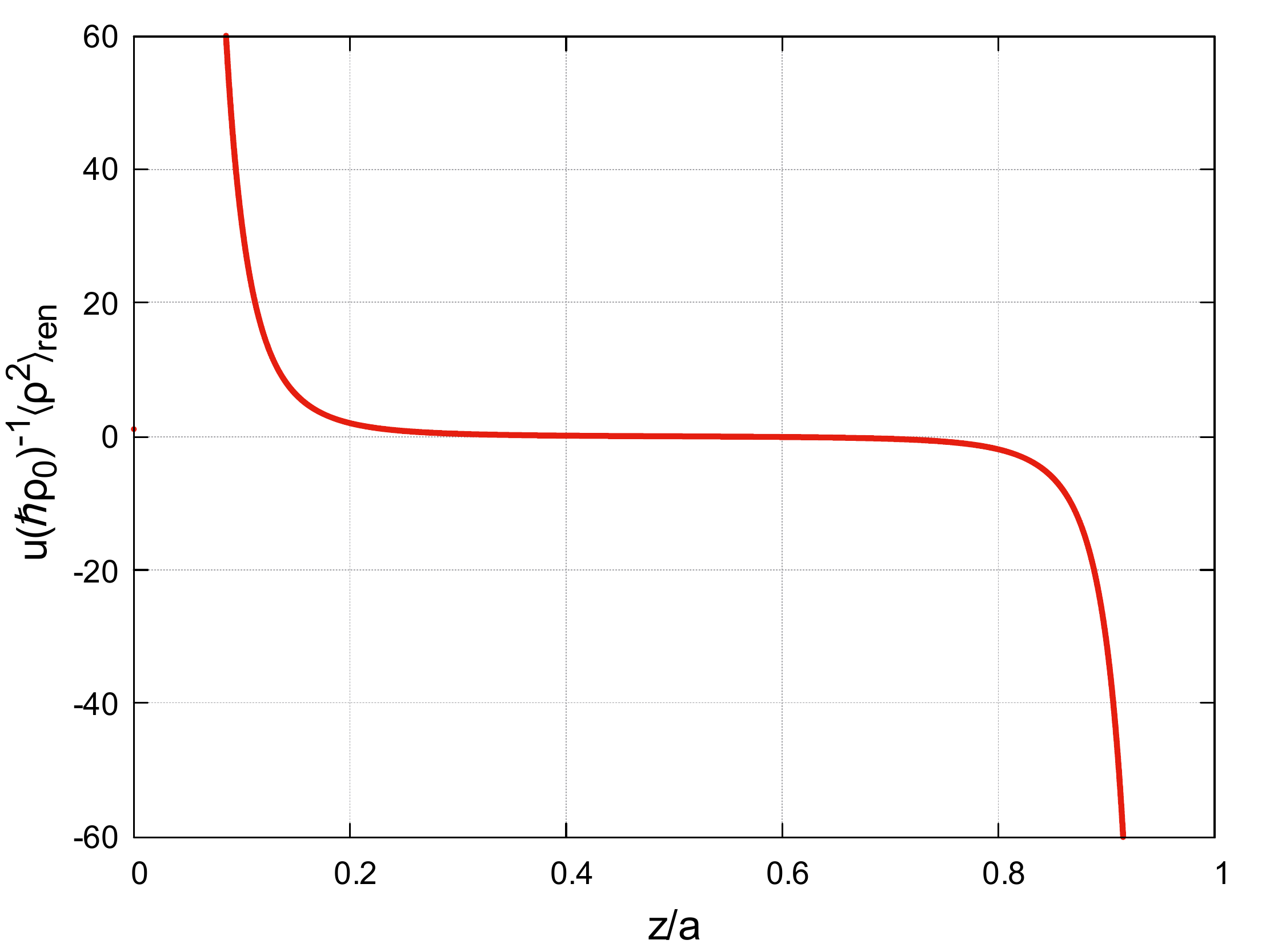}
\caption{Dimensionless renormalized mean squared density fluctuation, \eqref{3.16}, as a consequence of a mixed boundary condition, in terms of $z/a$.\small}
\label{figure2}
\end{center}
\end{figure}

In order to find the renormalized mean squared density fluctuation, $\langle\rho^2\rangle_{\rm ren}$, as a result of the implementation of the mixed boundary condition, we make use of Eqs. \eqref{2.15} and \eqref{3.12.1}. This gives
\begin{eqnarray}
\langle\rho^2\rangle_{\rm ren}=\frac{\hbar\rho_0}{32\pi^2 uz^4} - \frac{\hbar\rho_0}{2\pi^2 u}\int_{0}^{\infty}d\alpha\;\alpha^3\; \frac{\cosh(2\alpha z)-1}{3e^{2a\alpha}+3}.
\label{3.13}
\end{eqnarray}
The integral in $\alpha$ above can be divided into two integrals. The first one given by
\begin{eqnarray}
\int_{0}^{\infty}d\alpha\;\frac{\alpha^3}{3e^{2a\alpha}+3}=\frac{7\pi^4}{5760a^4}=\frac{7\zeta(4)}{64a^4},
\label{3.14}
\end{eqnarray}
where $\zeta(4)=\frac{\pi^4}{90}$ is the Riemann zeta function. On the other hand, the second integral can be computed as \cite{prudnikov1986integrals}
\begin{eqnarray}
\int_{0}^{\infty}d\alpha\;\alpha^3 \ \frac{\cosh(2\alpha z)}{3e^{2a\alpha}+3}&=&\frac{\partial^3}{\partial z^3}\int_{0}^{\infty}d\alpha\;\frac{\sinh(2\alpha z)}{8(3e^{2a\alpha}+3)}\nonumber\\
\ &=&\frac{\partial^3}{\partial z^3}\sum_{n=0}^{\infty}\int_{0}^{\infty}d\alpha\;\frac{1}{8(3e^{2a\alpha}+3)}\frac{(2\alpha z)^{2n+1}}{(2n+1)!}\nonumber\\
\ &=&\frac{1}{16z^4}-\frac{\pi^4}{96a^4}\frac{\cos(\pi z/a)}{\sin^4(\pi z/a)}(\cos^2(\pi z/a)+5),
\label{3.15}
\end{eqnarray}
where we have used the Taylor series expansion for the hyperbolic function $\sinh(x)$. Consequently, we have solved the integral in the second line first, performed the summation in $n$ next, and taken the derivatives with respect to $z$. Finally, with the integrals in Eqs. \eqref{3.14} and \eqref{3.15}, the mean squared density fluctuation \eqref{3.13} is written as
\begin{eqnarray}
\langle\rho^2\rangle_{\rm ren}=\frac{\hbar\rho_0\pi^2}{192ua^4}\left\{\frac{7}{60} + \frac{\cos(\pi z/a)}{\sin^4(\pi z/a)}(\cos^2(\pi z/a)+5)\right\}.
\label{3.16}
\end{eqnarray}
In Fig.~\ref{figure2} we have plotted the  dimensionless mean squared density fluctuation for the mixed boundary condition case given above, as a function of the dimensionless parameter $z/a$, where one of the planes passes the point $z=0$ and the other the point $z=a$. It is straightforward to see that, up to numerical constants, Eq. \eqref{3.16} diverges as $1/z^4$ near the plane at $z=0$ with Dirichlet and as $-1/(z-a)^4$ near the plane at $z=a$ with Neumann boundary conditions. This is evident in Fig.\ref{figure2}.

Let us now consider the mean squared density fluctuation for Dirichlet and Neumann boundary conditions. Thanks to the renormalized vacuum expectation value found in Eq. \eqref{2.15} with the two-point function in Eq. \eqref{3.04.1}, we end up with the final result
\begin{equation}
\langle\rho^2\rangle_{\rm ren}=-\frac{\hbar\rho_0\pi^2}{96ua^4}\left[\frac{1}{15}\pm\frac{3-2\sin^2(\pi z/a)}{\sin^4(\pi z/a)}\right],
\label{3.05}
\end{equation}
\begin{figure}[!htb]
\begin{center}
\includegraphics[width=0.5\textwidth]{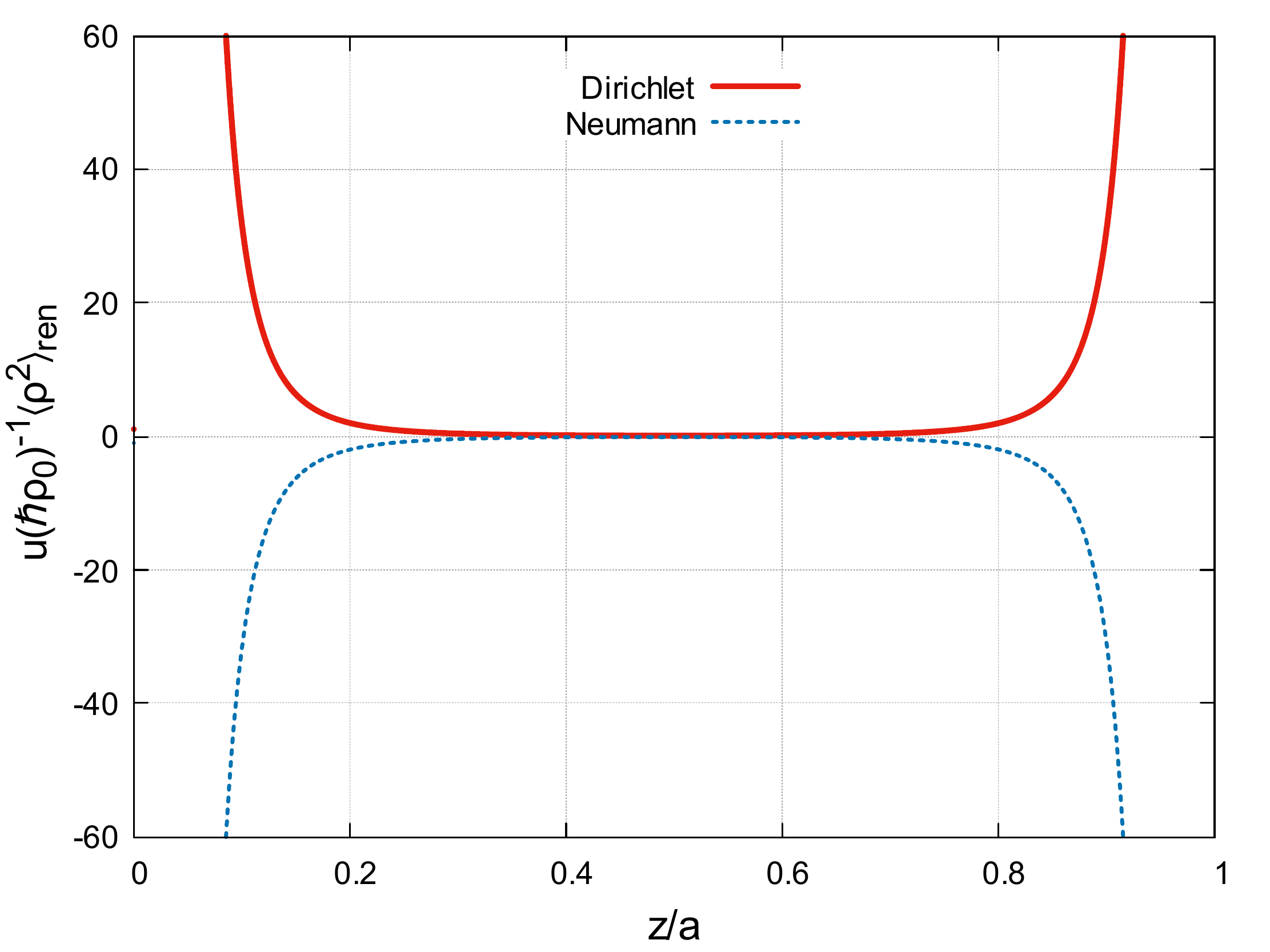}
\caption{Dimensionless renormalized mean squared density fluctuation, \eqref{3.05}, as a function of $z/a$.\small}
\label{figure3}
\end{center}
\end{figure}
where, once again, the minus and plus signs are associated with the Dirichlet and Neumann boundary conditions, respectively. As one can see, for the Neumann boundary condition case, the mean squared density fluctuation is always negative, which is in agreement with the result in \cite{ford2009fluid,ford2009phononic}. However, for the Dirichlet boundary condition, the second term is dominant, and as a consequence,
the mean squared density fluctuation is always positive, consistent with the result for the same boundary condition on one plane. We have plotted in Fig.\ref{figure3} the dimensionless mean squared density fluctuation \eqref{3.05} as a function of $z/a$ in both Dirichlet and Neumann cases. As in the mixed boundary case, we observe that, up to numerical constants, the renormalized mean squared density fluctuation diverges as $1/z^4$ in the limit $z\to 0$ and as $1/(z-a)^4$ in the limit $z\to a$ in the case of the Dirichlet boundary condition imposed on both planes. In contrast, considering the Neumann boundary condition, the mean squared density diverges with the opposite sign in both limits.
This behavior is shown in Fig.~\ref{figure3}.

It is interesting to note the transition from a positive to a negative value in  Fig.~\ref{figure3} for the mixed boundary condition case, where the renormalized mean square density fluctuation positively diverges on the plane under Dirichlet boundary condition and negatively diverges on the plane with Neumann boundary condition. This behavior is consistent with the result shown in Fig.~\ref{figure2}. Let us emphasize that both results for mixed and Dirichlet boundary conditions found, respectively, in Eqs. \eqref{3.16} and \eqref{3.05} have been obtained here for the first time, in the context of quantum vacuum fluctuations of phonons originating from liquid density fluctuation. The result in \eqref{3.05} for the mean squared density fluctuation under the Neumann boundary condition has been previously studied in Refs. \cite{ ford2009fluid,ford2009phononic}. It is clear that the study of phonons representing quantum excitations of a classical liquid's velocity potential, represented by a real scalar field, once subjected to boundary conditions such as mixed, Dirichlet and Neumann, as well as the nontrivial topology of a cosmic string, produce a nonzero renormalized mean squared density fluctuation, analogous to the Casimir effect in quantum field theory \cite{bordag2009advances}.


\section{Conclusion}\label{sec5}

This paper has investigated the effects in a classical liquid created by boundary conditions and the nontrivial topology of an ideal cosmic string in the mean squared density fluctuation. Considering the regime in which the wavelengths are much longer than the interatomic separation, we have assumed that the linear phonon dispersion relation is valid. 
Promoting the classical hydrodynamics quantities
to second quantized operators as well as relating the second quantized density fluctuations to the scalar velocity potential, the problem turns into the known Klein-Gordon for a massless scalar field with the light velocity replaced by the sound velocity.
In this scenario, phonons represent quantum excitations of a real massless scalar field associated with the classical liquid's velocity potential.

Furthermore, we have studied the quantized phonon field in the presence of the nontrivial topology of a cosmic string, as well as in the presence of Dirichlet, Neumann, and mixed boundary conditions in Minkowski spacetime. In the ideal (3+1)-dimensional cosmic string spacetime, with conical parameter $q>0$, we have also considered the phonon modes obeying a quasi-periodic condition, characterized by the parameter $\beta$. The case $q=1$ gives the results in the absence of the conical structure. Under the conditions mentioned above, the Klein-Gordon equation was solved, and the complete normalized solution is shown in \eqref{2.5}. The solution then was used to obtain an exact and analytical expression for the two-point function, that is, Eq. \eqref{2.11} which is in agreement with Ref \cite{klecio2020quantum}.

The closed expressions for the two-point function paved the path to obtaining the renormalized mean square density fluctuation of the liquid given in \eqref{2.16}. We have verified that in the case of $\beta = 0$, our results agree with previous results found in the literature \cite{ ford2009fluid,ford2009phononic}. Moreover, we have also shown that the renormalized mean squared density fluctuation is nonzero, even in the absence of the conical defect due to the quasi-periodicity condition. However, when $q=1$ and $\beta=0$, it vanishes, as expected. As it is usual in the cosmic string spacetime, as a result of it being ideal, the renormalized mean square density fluctuation diverges on the defect. A plot of the dimensionless renormalized mean squared density fluctuation in terms of the quasi-periodic parameter $\beta$ is shown in Fig.\ref{figure1}.

Additionally, we have investigated the phonon modes obeying mixed, Dirichlet, and Neumann boundary conditions, imposed on one and two parallel planes in Minkowski spacetime. We have, then, found an analytical form for the two-point functions in each case, that is, Eqs. \eqref{3.12.1} and \eqref{3.04.1}, as well as the renormalized mean square density fluctuation in Eqs. \eqref{3.16} and \eqref{3.05}. We have shown that the boundary conditions modify this physical quantity, as expected since the phonon quantum excitations in a liquid are analogous to the ones in quantum field theory in the presence of parallel plates with the same boundary conditions. We have obtained that the renormalized mean square density fluctuation is always positive for the planes with Dirichlet boundary condition and negative with Neumann boundary condition, regardless of the system's parameters. However, in the mixed boundary case, it can be negative or positive depending on the distance from the planes, characterized by the parameter $a$. This is consistent because the renormalized mean squared density fluctuation is positive when one considers only a plane with Dirichlet and negative when one considers only Neumann. In all these cases, the renormalized mean squared density fluctuation diverges on the planes. This is shown in the plots of Figs.~\ref{figure2} and \ref{figure3}.


{\acknowledgments}
The author K.E.L.F. is funded through a Ph.D. scholarship by the Brazilian agency CAPES. The author A.M. thanks financial support from the Brazilian agencies, CAPES and CNPq, grants 305893/2017-3 and 309368/2020-0, and also Universidade Federal de Pernambuco Edital Qualis A. The author H.F.S.M. is supported by the Brazilian agency CNPq under grants 305379/2017-8 and 311031/2020-0.

\bibliographystyle{JHEP}


\end{document}